\documentclass[useAMS,usenatbib]{mn2e}
\usepackage{amsmath}
\usepackage{amssymb}
\usepackage{graphicx}
\usepackage{epsfig}
\makeatletter{}\def\LCDM{\mbox{$\Lambda$CDM }}
\def\k{\mbox{$\,h$Mpc$^{-1}$}}

\def\Mpch{\mbox{$h^{-1}$Mpc}}
\def\Gpch{\mbox{$h^{-1}$Gpc}}

\def\M200{\mbox{$M_{\rm 200 }$}}

\def\R200{\mbox{$R_{\rm 200 }$}}

\def\V200{\mbox{$V_{\rm 200 }$}}

\def\Nbody{$N$-body}

\renewcommand{\vec}[1]{ {\bmath #1} }
\newcommand{\Ng}{\mbox{$N_{\rm g}$}}
\newcommand{\Np}{\mbox{$N_{\rm p}$}}
\newcommand{\Ns}{\mbox{$N_{\rm s}$}}
\newcommand{\lsim}{\mbox{${\,\hbox{\hbox{$ < $}\kern -0.8em \lower 1.0ex\hbox{$\sim$}}\,}$}}
\newcommand{\gsim}{\mbox{${\,\hbox{\hbox{$ > $}\kern -0.8em \lower 1.0ex\hbox{$\sim$}}\,}$}}

\def\beqn{\vspace{2mm}
\begin{eqnarray}} 
\def\eeqn{\vspaceg{2mm} 
\end{eqnarray}}

\newcommand{\be}{\begin{equation}}
\newcommand{\ee}{\end{equation}}
\newcommand{\ba}{\begin{eqnarray}}
\newcommand{\ea}{\end{eqnarray}}
\newcommand{\brr}{\begin{array}}
 
\newcommand{\err}{\end{array}}
\newcommand{\bc}{\begin{center}}
\newcommand{\ec}{\end{center}}

\title[Suppressing cosmic variance with paired-and-fixed cosmological
simulations]{Suppressing cosmic variance with paired-and-fixed cosmological
simulations: average properties and covariances of dark matter clustering statistics}

\makeatletter{}\author[A.~Klypin, F.~Prada \& J.~Byun]
  {Anatoly~Klypin$^{1}$, Francisco~Prada \& Joyce Byun 
   \vspace{0.2cm}\\ 
  $^1$Astronomy Department, New Mexico State University, Las Cruces, NM, USA\\
}

\author[Klypin, Prada \& Byun]{Anatoly~Klypin$^1$\thanks{E-mail: aklypin@nmsu.edu}, Francisco~Prada$^{2}$ \& Joyce~Byun$^{3}$ \\
 \vspace{-0.2cm}\\
$^{1}$ Astronomy Department, New Mexico State University, Las Cruces, NM, USA\\
$^2$ Instituto de Astrof\'{\i}sica de Andaluc\'{\i}a (CSIC), Glorieta de 
     la Astronom\'{\i}a, E-18080 Granada, Spain \\
$^3$ D\'epartement de Physique Th\'eorique, University of Geneva, Geneva, Switzerland \\
\\
}

\begin{document}
\maketitle
\label{firstpage}
\begin{abstract}
  Making cosmological inferences from the observed galaxy clustering
  requires accurate predictions for the mean clustering statistics and
  their covariances. Those are affected by cosmic variance -- the
  statistical noise due to the finite number of harmonics.  The cosmic
  variance can be suppressed by fixing the amplitudes of the harmonics
  instead of drawing them from a Gaussian distribution predicted by
  the inflation models. Initial realisations also can be generated in
  pairs with $180^{\circ}$ flipped phases to further reduce the
  variance. Here, we compare the consequences of using
  paired-and-fixed vs Gaussian initial conditions on the average dark
  matter clustering and covariance matrices predicted from $N$-body
  simulations.  As in previous studies, we find no measurable
  differences between paired-and-fixed and Gaussian simulations for
  the average density distribution function, power spectrum and
  bispectrum.  Yet, the covariances from paired-and-fixed simulations
  are suppressed in a complicated scale- and redshift-dependent way.
  The situation is particularly problematic on the scales of Baryon
  Acoustic Oscillations where the covariance matrix of the power
  spectrum is lower by only $\sim 20\%$ compared to the Gaussian
  realisations, implying that there is not much of a reduction of the
  cosmic variance.  The non-trivial suppression, combined with the
  fact that paired-and-fixed covariances are noisier than from
  Gaussian simulations, suggests that there is no path towards
  obtaining accurate covariance matrices from paired-and-fixed
  simulations.  Because the covariances are crucial for the
  observational estimates of galaxy clustering statistics and
  cosmological parameters, paired-and-fixed simulations, though useful
  for some applications, cannot be used for the production of mock
  galaxy catalogs.
\end{abstract}

\begin{keywords}
cosmology: Large-scale structure of Universe - dark matter - methods: numerical
\end{keywords}

\makeatletter{}\section{Introduction}

Large galaxy redshift surveys yield vital knowledge
on numerous properties of the large-scale structure of the universe, 
many of the key cosmological parameters and the nature of dark matter. While the signal---whatever
statistics are used---is measured using observational data, its
uncertainties and errors are typically obtained by theoretical
modelling and understanding the phenomena involved
\citep[e.g.,][]{Mandelbaum2013,Sergio2016,DES2018,Uitert2018}. Cosmological
simulations play a crucial role in the process of estimating the errors.

More specifically, cosmological simulations generate two types of
results: i) the average statistics -- such as the power spectrum,
correlation function, or weak-lensing signal -- for the adopted cosmological
parameters, and ii) covariance matrices, i.e., the correlation of the measured statistics at
different bins. For example, this can be the
correlation of power spectra values at different wavenumbers or the
correlations of values of the correlation function at different radii. 

The average properties of clustering statistics are important on their own. They provide a preliminary test of
whether a given cosmological model can reproduce the observations. If
it clearly cannot, there is no need to perform a detailed analysis constraining the particular cosmological model.
Average properties themselves are difficult
to estimate. This requires a procedure to connect dark matter with galaxies, which
is one of the fundamental problems of modern cosmology
\citep[e.g.,][]{Somerville2015,Wechsler2018}. There are different
methods and techniques for how to do this in the context of large
cosmological surveys \citep[e.g.,][]{Somerville1999,Conroy2006,
  Sebstian2011,Sergio2016,Monaco2016}.

However, once we know how to estimate the average statistics, we face
an even more difficult challenge: we need to produce many realisations
of the statistics to estimate the covariance matrix which describes
the errors of the observations. The number of realisations depends on
the particular statistics that are used, but typically one needs to
run many thousands of simulations to obtain accurate estimates of
error bars and covariance matrices \citep[e.g. see][and references
therein]{GLAM}. This puts significant stress on theoretical
predictions, and practically discards some computational methods such
as direct hydro-dynamical simulations of galaxy formation or
high-resolution $N$-body simulations. They will be too expensive to
run for thousands of realisations.

In the last few years, new techniques have been developed to address
the need for massive production of mock galaxy samples. These include
$N$-body codes such as COLA \citep{Tassev2015,Koda2016} and GLAM
\citep{GLAM} that are very fast and have sufficient resolution for
some observational statistics. There are also approximate methods that
lack predictive power, but can be used for estimates of covariances
\citep{Patchy2016,Monaco2016,EZMOCKS2018,Lippich2019}.

A different approach has been recently proposed by
\citet{AnguloPontzen}. In order to reduce the scatter between
different realisations, i.e., the effects of cosmic variance, they
suggest fixing the amplitudes of the Fourier harmonics. Instead of
having purely Gaussian initial conditions, as typically predicted by
inflation models \citep{Mukhanov1992,Inflation2004}, the amplitudes of
the harmonics are fixed to have the same magnitude as the ensemble
average power spectrum. In addition, the realisations can be run in
pairs with the phases flipped by $180\,^{\circ}$ to further reduce the
noise \citep{Pontzen2016}. The real fluctuations which originated
during inflation should not have fixed amplitudes, meaning that the
paired-and-fixed method is a trick to reduce the noise. However, this
method can be quite useful. Its success comes from the fact that in
the linear regime the fixed-amplitude density perturbation field
$\delta\rho(\vec x,t)$ has a Gaussian distribution with the
correlation function being the same as for normal perturbations, i.e.,
$\langle\delta\rho(\vec x,t)\delta\rho(\vec x+ \vec
x^\prime,t)\rangle$.
This implies, for example, that it has the same statistics for the
high-$\sigma$ peaks responsible for the formation of the most massive
halos.  It has the same mixture of long and short wavelengths (as
manifested by the power spectrum), and thus the same sequence of
growing non-linear structures.

The paired-and-fixed method has been implemented and studied in a
number of publications. \citet{AnguloPontzen} found that the dark matter power spectrum from a single pair of fixed simulations reproduces the average power spectrum nearly as well as an ensemble of 300 standard Gaussian simulations at $z=1$, even on deeply non-linear scales, $k\approx 1\k$. Similarly good agreement was found for the mass function, bispectrum
and correlation function. However, because only one pair of fixed simulations was used for the comparison, the residual r.m.s. errors of the statistics from the paired-and-fixed method, relative to the standard method, were noisy. For example, the r.m.s. deviations for the power spectrum were
very small on large scales ($k\lsim 0.02\k$), but they became increasingly noisy on
smaller scales.  On the other hand, the r.m.s. deviations of the mass function did not become smaller by using the paired-and-fixed  simulations.
The paired-and-fixed method thus appears to be a powerful tool for quickly achieving small r.m.s. errors for some statistics, but more extensive analyses of the
r.m.s. residuals and covariances are clearly needed.
More realisations of paired-and-fixed simulations could be used to examine the average r.m.s. errors, and comparison with different simulation codes could confirm the robustness of the method.

\citet{PairedStatistics} used 100 realisations of both Gaussian and
paired-and-fixed simulations using the {\sc Gadget} code with $512^3$
particles in a 1\,\Gpch\, box. Because of the high force resolution of
the {\sc Gadget} simulations, the analysis of the power spectrum was
done up to $k\sim 1\k$. As in \citet{AnguloPontzen}, they also found a
dramatic suppression in the scatter of the matter power spectrum at
long wavelengths: a factor of $10^3$ in the r.m.s. at $k=0.01\k$.  At
smaller scales, the r.m.s. tends towards the same result as for
Gaussian simulations, but upon closer inspection the r.m.s. errors for
the matter power spectrum are systematically smaller.
\citet{PairedStatistics} also studied the density probability
distribution function (PDF) for the relatively large cell size of
$\sim 8\,\Mpch$, which probed the density field up to modest
over-densities, $\delta\rho/\rho< 50$. Overall, their main conclusion
was that the scatter (cosmic variance) can be strongly suppressed for
the power spectrum on large scales, but not as much for the mass
function or the power spectrum on small scales. The relatively small
number of realisations translated into noisy r.m.s. values and did not
allow for an analysis of the covariances.

One of the main goals of large-volume cosmological simulations is to
estimate the scatter and covariances of statistics observed by large
galaxy surveys. In that respect the main advantage of the
paired-and-fixed simulations---reduced scatter---begins to appear
problematic. Indeed, the realisation-to-realisation scatter, such as
the scatter in the power spectrum or bispectrum at a fixed wavenumber,
is simply the diagonal component of the covariance matrix. If it is
\textit{reduced}, then the covariance matrix will be incorrect. To
date the scatter has been measured with large errors, and there has
been no exploration of the effect on full covariance matrices.

With 400 realisations of a $1\,\Gpch$ simulation box \citet{UNIT} had
enough data to study the scatter of the power spectrum in more detail
than \citet{PairedStatistics}. They found that relative to the
Gaussian simulations, the scatter in paired-and-fixed simulations was
scale-dependent and approaching unity at $k\gsim 0.2\k$ and
$z=1$. However, at these small scales the r.m.s. estimates were too
noisy to determine whether the ratio of Gaussian to paired-and-fixed
r.m.s. was indeed constant.  \citet{UNIT} also argued that because the
non-diagonal covariance matrix scales as $1/V$, where $V$ is the
simulation volume, the full covariance matrix should be smaller
compared to the Gaussian simulations by a factor of $1/V$. This would
imply that a single pair of fixed simulations is effectively measuring
the covariance matrix of a large ensemble of Gaussian simulations.
However, this reasoning was not substantiated by an analysis of the
simulations. It also contradicts the authors' own results which show
that the diagonal components of the covariance matrix depend on the
wavenumber $k$, and thus cannot simply be re-scaled by $1/V$ to obtain
the same covariance matrix as an ensemble of Gaussian simulations.
The results presented in this paper contradict the \citet{UNIT} claim
that paired-and-fixed covariance matrices can be scaled with the
simulation volume.

The main goal of our paper is to study in detail, using thousands of $N$-body simulations,
 the r.m.s. scatter and covariance matrices measured from paired-and-fixed simulations. How do
they depend on scale and redshift? Is this dependence a simple one which can be rescaled to recover the covariances of the true Gaussian
simulations? To answer these questions we made a very large number ($\sim 4000$) of
realisations with Gaussian and paired-and-fixed initial
conditions using the GLAM code. This allows us to study the diagonal and off-diagonal
covariances of the power spectrum and the diagonal covariances of the
bispectrum.

Section~\ref{sec:FlipFlop} provides some background on
paired-and-fixed, as well as Gaussian, initial conditions.
In Section~\ref{sec:simulations} we give the
details of our GLAM simulations. The average power spectrum and density
distribution function results are given in Section~\ref{sec:powersp}. In
Section~\ref{sec:covmatrix} we discuss the results on the covariance matrix
of the power spectrum. The impact on the bispectrum and its scatter is presented in Section~\ref{sec:bispectrum}.
We end with a discussion of our main conclusions in Section~\ref{sec:conclusion}.

\makeatletter{}\section{Paired-and-fixed simulations} 
\label{sec:FlipFlop}
The adopted computational domain is a cubic box of size $L$ and volume $V=L^3$. In this case the
fundamental mode $k_f= 2\pi/L$ of this domain defines the discreteness in
Fourier space: $\vec{k} =(k_x,k_y,k_z)= (i,j,k)k_f$, where $(i,j,k)$ are integer numbers.
 The density contrast in the domain can be expanded into the Fourier series as follows
\begin{equation}
\delta(\vec{x},t)\equiv \frac{\delta\rho}{\rho} =\sum_{\vec k}\delta_{\vec k}e^{i\vec k\vec x}.
\label{eq:delta}
\end{equation}
In the linear regime the amplitudes of the harmonics $\delta_{\vec k}$ are
uncorrelated random numbers that must obey two conditions. First, because
$\delta(\vec{x})$ are real numbers,
$\delta_{-\vec k}=\delta_{\vec k}^*$. Second, in the linear regime the curl of the velocity field must
be equal to zero.\footnote{This is because the vorticity 
represents a
decaying mode of the fluctuations in the linear regime. The
condition that $\vec{\nabla}\times\vec{v}=0$ imposes a certain
combination of signs of the harmonics. It is implemented in the initial
conditions of the cosmological simulation codes, but for clarity we ignore it here.}
  The condition that $\delta_{-\vec k}=\delta_{\vec k}^*$
 implies that only half of the harmonics in Fourier space are
independent. We can explicitly use this by writing the summation given in eq.~(\ref{eq:delta}) over half
of the Fourier space \citep[][eq. 9.10]{Binney}:

\begin{eqnarray}
\delta(\vec{x},t) &=&{\sum} ^\prime\delta_{\vec k}e^{i\vec k\vec x}+\delta_{-\vec k}e^{-i\vec k\vec x} \\
     &=& \alpha {\sum} ^{\prime}\sqrt{P(|\vec k|)}[A\cos(\vec k\vec x)+B\sin(\vec k\vec x)] \label{eq:delta3}\\
     &=& \alpha  {\sum} ^{\prime}\sqrt{P(|\vec k|)}C\cos(\vec k\vec x +\phi)\label{eq:delta4}.
\end{eqnarray}
Here $\alpha$ is a normalization factor, $P(|\vec k|)$ is the initial power spectrum for a given cosmology, and the parameters 
 $A$ and $B$ are random numbers following a Gaussian distribution
with zero mean and dispersion unity. In eq.~(\ref{eq:delta4}) the parameter
$C$ is a random number following a chi distribution with two degrees of
freedom, $C=\sqrt{A^2+B^2}$, which is also known as the Rayleigh
distribution. The parameter $\phi$ is a random number that is uniformly distributed between 0 and $\pi$.

Both forms for the density spectrum, eq.~(\ref{eq:delta3}) and
eq.~(\ref{eq:delta4}), are widely used for setting up the initial conditions.
We prefer to use eq.~(\ref{eq:delta3}) because it is easy to implement
for real-to-real FFT routines, which require only half as much memory to be allocated
compared to complex-to-complex FFTs. Otherwise they are mathematically identical.

In order to generate paired initial conditions, one simply adds $\pi$
to the phase $\phi$ in eq.~(\ref{eq:delta4}) or changes the signs of $A$
and $B$ in eq.~(\ref{eq:delta3}). In both cases the sum of paired
realisations gives $\delta(\vec{x})+\delta_{\rm paired}(\vec{x})=0$.
This setup with the Gaussian or Rayleigh random numbers sets the initial conditions for perturbations generated during inflation. 

For fixed amplitude simulations one assumes that $A$ and $B$ are
random numbers, but their distribution is peculiar: they are either
one or minus one, i.e. $A,B =(-1,1)$.  By design, the sum of squares of the amplitudes of the harmonics for
each narrow bin $(k,k+\Delta k)$ in Fourier space returns exactly the input
power spectrum $P(k)$. This is not the case for truly Gaussian initial conditions
where the power spectrum of a given realisation has the following r.m.s. deviations:

\begin{equation}
\frac{\Delta P}{P} = \left(\frac{2}{N_h}\right)^{1/2},\quad N_h=\frac{4\pi k^2\Delta k} {k_f^3},
\label{eq:Gauss}
\end{equation}
where $k_f$ and $N_h$ are the fundamental harmonic and the number of harmonics, respectively. 

In spite of the fact that fixed amplitude initial conditions do not
have the correct distribution of the amplitudes in phase space, its
density field $\delta(\vec{x})$ has the same distribution as in the
purely Gaussian initial conditions: it is a Gaussian field with the same
correlation function by virtue of the central limit theorem. The
sum of random variables with the same distribution---in our case the $A$
and $B$ variables in eq.~(\ref{eq:delta3})---has a Gaussian distribution.

In the same way, one expects that any statistic that is a convolution
of many Fourier components will have the same average property in both
fixed and Gaussian simulations. For example, the
correlation function is a convolution of the power spectrum:

\begin{equation}
\xi(r)  = \frac{1}{2\pi^2}\int_0^\infty k^2dkP(k)\frac{sin(kr)}{kr}.
\label{eq:xi}
\end{equation}
Thus, it must be the same for fixed and Gaussian simulations. The same holds
for the mass function.

However, one expects differences for the second-order statistics: the
r.m.s. scatter and covariance matrices. Indeed, there is no scatter of
the power spectrum or the correlation function for fixed
simulations. This likely means that other statistics, such the r.m.s. of the bispectrum,
may also be affected.

The situation becomes more complicated in the non-linear regime. The
fixed amplitude perturbations still produce random fields: the
combination of ``+'' and ``--'' signs for the parameters $A$ and $B$ varies from
realisation to realisation. Non-linear interactions affect these
random fields in a complicated way. As we will see later,
non-linearities typically amplify the random nature of the
perturbations resulting in a substantial increase of the scatter and
covariances.

Considering these potential effects and the existing results in the literature, in this paper we have two main goals: i) to understand the non-linear
effects and accuracy of paired-and-fixed simulations by
studying the power spectrum and high-mass end of the density distribution
function, and ii) to examine the scatter and covariance matrices of the power spectrum and bispectrum from paired-and-fixed simulations.

\makeatletter{}\begin{table*}
 \begin{minipage}{16.cm}
\caption{Numerical  parameters of the GLAM simulations used in this work.
  The columns give the simulation identifier, 
  the size $L$ of the simulated box in $h^{-1}\,{\rm Mpc}$,
  the number of particles $\Np^3$, 
  the mass per particle $m_p$ in units of $h^{-1}\,M_\odot$, the mesh size $\Ng^3$,
  the  gravitational softening length $\epsilon$ in units of $h^{-1}\,{\rm Mpc}$,
  the number of realisations $N_r$ and the total volume in $h^{-1}\,{\rm Gpc}^3$.}
\begin{tabular}{ l | c | c | c |  c|  c | r | r |l }
\hline  
Simulation & $L^3$ &  $\Np^3$  & $m_p$                    & $\Ng^3$  & $\epsilon$   & $N_r$ & $V_{\rm tot}$ & Initial conditions
\tabularnewline
  \hline 
1GpcGauss  & 1000$^3$    & 1000$^3$ & $8.5\times 10^{10}$   & 2000$^3$ & 0.50 & 1200 & 1200 & Gaussian fluctuations \\
1GpcFixed  & 1000$^3$    & 1000$^3$ & $8.5\times 10^{10}$   & 2000$^3$ & 0.50 & 600 & 600   & Fixed amplitude fluctuations   \\
1GpcPairedFixed & 1000$^3$    & 1000$^3$ & $8.5\times 10^{10}$   & 2000$^3$ & 0.50 & 600 & 600   & Paired simulations for 1GpcFixed \\
3GpcGauss  & 3000$^3$    & 1000$^3$ & $2.3\times 10^{12}$   & 2000$^3$ & 1.50 & 1300 & 35000 & Gaussian fluctuations \\
3GpcFixed  & 3000$^3$    & 1000$^3$ & $2.3\times 10^{12}$   & 2000$^3$ & 1.50 & 25 & 675    & Fixed amplitude fluctuations \\
3GpcPairedFixed  & 3000$^3$    & 1000$^3$ & $2.3\times 10^{12}$   & 2000$^3$ & 1.50 & 25 & 675   & Paired simulations for 3GpcFixed \\
C1.2       & 1200$^3$    & 2000$^3$ & $1.8\times 10^{10}$   & 5000$^3$ & 0.24 & 22 &  38   & Gaussian fluctuations 
\tabularnewline
\hline          
\end{tabular}
\label{table:simtable}
\vspace{-5mm}
\end{minipage}
\end{table*}

\makeatletter{}\section{Simulations} 
\label{sec:simulations}

We use the Parallel Particle-Mesh $N$-body code {\sc GLAM} \citep{GLAM} 
that provides us with a tool to quickly generate
a large number of \Nbody\ cosmological simulations with reasonable
speed and acceptable numerical resolution. For a particular cosmological model and
initial conditions, GLAM generates the density field,
including peculiar velocities.  The code uses a regularly spaced
three-dimensional mesh of size $N_{\rm g}^3$ that covers the cubic
domain $L^3$ of a simulation box using $N_{\rm p}^3$
particles. The size of a cell $\epsilon =L/N_{\rm g}$ and the mass of
each particle $m_{\rm p}$ define the force and mass resolutions,
respectively.

The number of time-steps $\Ns$ is proportional to the computational
cost of the simulation. This is why reducing the number of steps is
important for producing a large set of realisations. \citet{QPM} and
\citet{Koda2016} use just $\sim 10$ time-steps for their QPM
and COLA simulations. \citet{Feng2016} and \citet{Izard2015} advocate using
$\Ns\approx 40$ steps for Fast-PM and ICE-COLA.

Because our main goal is to produce simulations requiring minimal
corrections to the local density and peculiar velocities, we use
$\Ns = 136$ time-steps in our  GLAM simulations. This
number of steps also removes the need to split particle displacements
into quasi-linear ones and the deviations from quasi-linear
predictions. In this way we greatly reduce the complexity of the
code and increase its speed, while also substantially reducing the memory
requirements. See  \cite{GLAM} for more details.

We use a very large number of GLAM simulations---about 4,000---to study
different aspects of dark matter clustering statistics in the flat $\LCDM$ Planck cosmology with parameters
$\sigma_8=0.828$, $\Omega_\Lambda=0.693$, $\Omega_m=0.307$,
$\Omega_b=0.048$ and $h=0.678$. The numerical parameters of our
simulations are presented in Table~\ref{table:simtable}. All of the
simulations were started at an initial redshift $z_{\rm init}=100$ using
the Zeldovich approximation.

When generating the initial conditions for fixed amplitude simulations, we
use the same code that generates true Gaussian simulations. However,
the amplitudes of the Fourier harmonics are replaced by either one or minus
one depending on the sign of the random Gaussian number used in normal
simulations. To generate a paired realisation, we flip the phases by $180\,^{\circ}$.

The analysis of paired and fixed simulations is done by first averaging
the results within each pair and then getting the statistics of the averages.
This procedure reduces the scatter by $\sqrt{2}$ even for completely
independent realisations. In order to compensate for this effect we
scale up the scatter by $\sqrt{2}$, which effectively provides the scatter
per realisation.

\makeatletter{}\section{Power Spectrum and Density Distribution Function}
\label{sec:powersp}

\makeatletter{}\begin{figure}
\centering
\includegraphics[width=0.475\textwidth]
{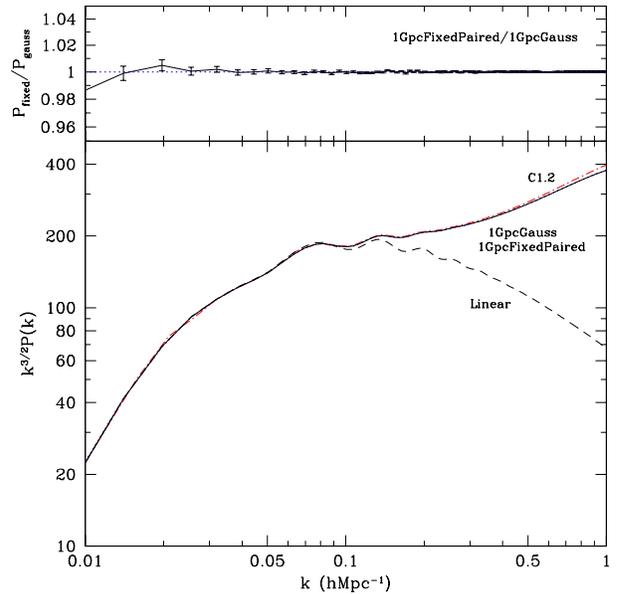}

\caption{{\it Bottom panel:} Power spectra of dark matter at redshift
  $z=0$ scaled by the factor $k^{3/2}$ to highlight the BAO region at
  $k=(0.05-0.30)\,h{\rm Mpc}^{-1}$. The long-dashed curve shows the
  linear theory.  The results for the $1\,\Gpch^3$ paired-and-fixed (full) and purely Gaussian
  (short-dashed) simulations are so close that one cannot
  distinguish between the two at very long ($\sim 1\,\Gpch$) or short
  ($\sim 1\,\Mpch$) wavelengths in the deeply non-linear regime. {\it Top panel:}
  Ratio of the average power spectra with error bars from Gaussian simulations. There are no detectable
  differences between paired-and-fixed and Gaussian simulations.
}
\label{fig:Power}
\end{figure}

Figure~\ref{fig:Power} shows our results for the average power spectra at
$z=0$ for 1200 realisations with true Gaussian initial conditions (1GpcGauss, short-dashed) and 
another 1200 realisations of paired-and-fixed fluctuations (1GpcPairedFixed, full curve). 
For comparison we also show the linear power spectrum (long-dashed curve) and the
average power spectrum from 22 realisations of a higher resolution simulation  run
with Gaussian perturbations (C1.2, dot-dashed).

A more detailed analysis of the convergence of the power spectrum from GLAM
simulations was presented in \citet{GLAM}. For the numerical resolution of the
1GpcGauss simulations the power spectrum has an error of $\lsim 5$ per cent
at $k=1\,\k$ and $\lsim 1$ per cent at $k=0.6\,\k$. The simulations with
$3\Gpch$ boxes converge with $\lsim 2$ per cent error at $k=0.3\,\k$.
However, here we are interested in the relative differences between
Gaussian and paired-and-fixed simulations with the same resolution. Therefore, we can measure the differences for somewhat larger wavenumbers than the formal convergence limits. Indeed, as Figure~\ref{fig:Power} indicates, there are no
measurable differences of the average power spectrum between Gaussian
and paired-and-fixed simulations. These results are quite remarkable: for
$k>0.03\k$ the differences are less than $0.1$ per cent.

\begin{figure}
\centering
\includegraphics[width=0.48\textwidth]
{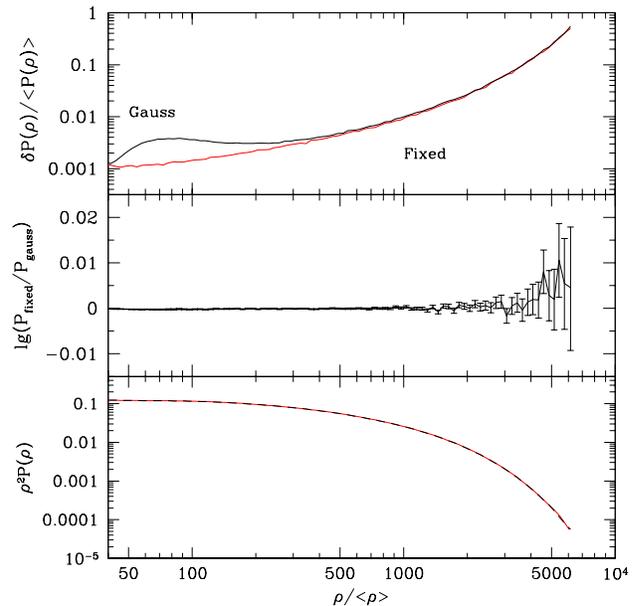}
\caption{Density distribution function. {\it Bottom panel:} $P(\rho)$
  scaled with $\rho^2$ to reduce the dynamical range. {\it Middle
    panel:} The ratio of the average PDF of paired-and-fixed
  simulations to those of true Gaussian simulations. {\it Top panel:} The
  r.m.s. scatter of the PDF of individual realisations relative to the
  ensemble average PDF. There are no detectable differences in the average
  PDFs even at very high densities with 
  $\rho/\langle\rho\rangle \approx 6000$, where there are only 2-3 such cells in each
  realisation. The paired-and-fixed simulations decrease the noise of the
  PDF by a factor of $\sim 1.5$ at large densities near $\rho/\langle\rho\rangle \sim 70$, but this merely suppresses a scatter that was very small to begin with. The noise is orders of magnitude larger at the high-density end of the PDF, where there is no reduction in the noise by using paired-and-fixed simulations.}
\label{fig:PDF}
\end{figure}

We also study the density distribution function $P(\rho) = dN/d\rho$, i.e., the 
probability that a cell has density $\rho/\langle\rho\rangle$. For
a relatively large cell size of $8\,\Mpch$ this statistic was studied by
\citet{PairedStatistics}. Here, we are interested in much smaller cell
sizes of $0.5\,\Mpch$ to probe the tail of $P(\rho)$ with extremely high densities
$\rho/\langle\rho\rangle\sim 10^4$. These densities correspond to the interiors of massive dark matter halos. Do paired-and-fixed simulations have
defects at these large densities? Do they reduce the scatter in the PDF? As Figure~\ref{fig:PDF} shows, 
 the answer to both questions is {\it no}. Just as with the average power spectrum, there are no differences
between Gaussian and paired-and-fixed simulations. However,
the paired-and-fixed simulations do not suppress the noise in the PDF
by a substantial amount either. In that regard, one does not gain an advantage
by using paired-and-fixed simulations.

\makeatletter{}

\makeatletter{}\section{Covariance matrix of the power spectrum}
\label{sec:covmatrix}

The covariance matrix $C(k,k^\prime)$ is the second-order statistic
of the power spectrum. The power spectrum covariance and its cousin, the
covariance of the correlation function, play a key role in 
estimating the accuracy of the power spectrum measured from large galaxy surveys. 
Furthermore, the inverse
covariance matrices are used to estimate the cosmological parameters
inferred from these measurements
\citep[e.g.,][]{Anderson2012,Sanchez2012,Dodelson2013,Percival2014}. The
power spectrum covariance matrix measures the degree of non-linearity
and mode coupling of waves with different wavenumbers. As such it is
an interesting entity on its own.

\makeatletter{}\begin{figure}
\centering
\includegraphics[width=0.490\textwidth]{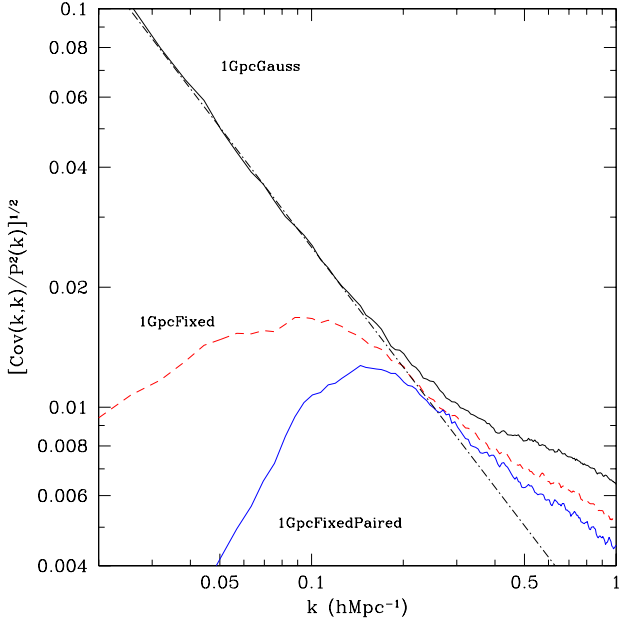}

\caption{Diagonal components of the covariance matrix $C(k,k)$ of
  the dark matter power spectrum at redshift $z=0$. Results are normalised to the signal
  $P^2(k)$ as indicated in the plot. For true Gaussian simulations (1GpcGauss, top
  full curve) the covariance matrix at $k\lsim 0.1\,h{\rm Mpc}^{-1}$
  is dominated by the shot-noise due to the finite number of harmonics
  in a given bin (dot-dashed curve). At larger wavenumbers the covariance
  matrix is dominated by non-linear effects. At small wavenumbers
  $k\lsim 0.1\,h{\rm Mpc}^{-1}$ the fixed (1GpcFixed, dashed line) and paired-and-fixed 
  (1GpcPairedFixed, bottom full curve) simulations 
  produce much smaller r.m.s. fluctuations of $P(k)$ per individual
  realisation. The r.m.s of 1GpcFixed is 20 times smaller at
  $k=0.02\,h{\rm Mpc}^{-1}$, implying that one would need 400 times
  more true Gaussian simulations to reach the same level of errors in the
  average power spectrum.  However, the difference in the diagonal
  components of the covariance matrix decreases dramatically with
  scale. For example, at $k=0.5\,h{\rm Mpc}^{-1}$ the r.m.s. fluctuation
  of the 1GpcFixedPaired power spectrum is just 20 per cent lower as compared to the Gaussian simulations.}
\label{fig:DiagCovMatrix}
\end{figure}

\makeatletter{}\begin{figure}
\centering
\includegraphics[width=0.480\textwidth]{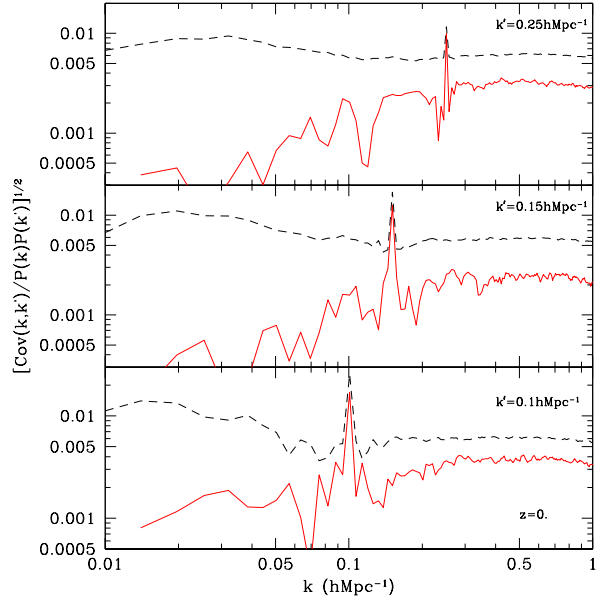}

\caption{Covariance matrix $C(k,k\prime)$ normalised to the signal $P(k)P(k^\prime)$ as
  indicated in the plot. The plot shows three cuts of the non-diagonal
  components at redshift $z=0$ (from bottom to top):
  $k^\prime=0.10, 0.15, 0.25\,h{\rm Mpc}^{-1}$. Full curves show the
  results for the 1GpcPairedFixed simulations. 1GpcGauss simulations are
  represented by dashed curves. As we showed in Figure~\ref{fig:DiagCovMatrix} for the diagonal components, the
  non-diagonal terms here show a substantial suppression at small scales and
  smaller differences at larger wavelengths. The differences are scale-dependent, indicating that it would be difficult to build a mock galaxy catalog from paired-and-fixed simulations that would
  faithfully recover the true Gaussian covariance matrix.}
\label{fig:CovMatrix}
\end{figure}

\makeatletter{}\begin{figure}
\centering
\includegraphics[width=0.480\textwidth]
{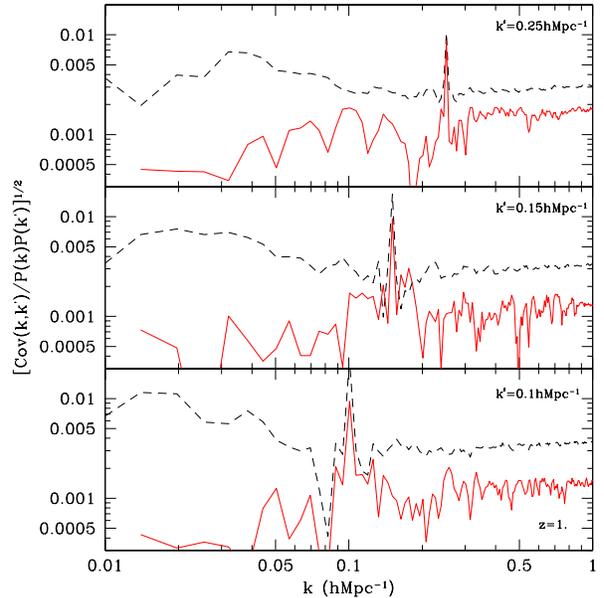}

\caption{The same as in Figure~\ref{fig:CovMatrix}, but at redshift
  $z=1$. Because the amplitude of fluctuations is smaller at higher redshifts, the covariances at
  $z=1$ are significantly noisier compared with $z=0$.}
\label{fig:CovMatrixz}
\end{figure}

The covariance matrix $C(k,k^\prime)$ is defined as a reduced cross
product of power spectra at different wavenumbers in the same
realisation averaged over different realisations, i.e.
\begin{equation}
  C(k,k^\prime) \equiv  \langle P(k)P(k^\prime)\rangle - 
                  \langle P(k)\rangle\langle P(k^\prime)\rangle.
\label{eq:Cov}
\end{equation}
The diagonal and non-diagonal components of the covariance matrix typically have very different magnitudes and evolve differently with
redshift. The diagonal elements are larger than the off-diagonal ones,
but there are many more off-diagonal elements, making them cumulatively
important \citep{Taylor2013,Percival2014,OConnell2016}. Off-diagonal
elements are solely due to non-linear clustering effects: in a
statistical sense the off-diagonal elements are equal to zero in the
linear regime. The diagonal component $C(k,k)$ can be written as a
sum of Gaussian fluctuations, due to the finite number of harmonics
in a given bin, and a term that is due to the non-linear growth of fluctuations, i.e.,
\begin{equation}
C(k,k) \equiv C_{\rm Gauss}(k) + C_{\rm non}(k,k),
\end{equation}
where the Gaussian term depends on the amplitude of the power spectrum
$P(k)$ and on the number of harmonics $N_h$ as
\begin{equation}
C_{\rm Gauss}(k) = \frac{2}{N_h}P^2(k),\quad N_h=\frac{4\pi k^2\Delta k}
                                                      {\left(2\pi/L\right)^3}.
\label{eq:GaussCov}
\end{equation}
Note that for a fixed bin width $\Delta k$ the number of harmonics is proportional to the computational volume, $N_h \propto L^3$,
and thus the amplitude of the Gaussian term scales as $C_{\rm Gauss}(k)\propto 1/L^3$ \citep[see][]{GLAM}.
 
Figure~\ref{fig:DiagCovMatrix} presents the diagonal components of the
covariance matrix.  As expected, the true Gaussian simulations closely
follow $C_{\rm Gauss}(k)$ at small wavenumbers $k\lsim 0.1\k$,
while the paired-and-fixed simulations show a dramatic reduction in the
scatter of the power spectrum $P(k)$. However, the situation changes in the non-linear regime $k\gsim 0.2\k$. There, the covariance matrix $C(k,k)$ of
the paired-and-fixed simulations increases substantially and becomes
only $\sim 20$ per cent smaller than the Gaussian simulations. In
addition, the ratio of the covariance matrices is $k$-dependent. 

The non-diagonal components $C(k,k^\prime)$ of the covariance matrix
show a similarly complicated trend as demonstrated by
Figure~\ref{fig:CovMatrix} and
Figure~\ref{fig:CovMatrixz} for $z=0$ and $z=1$, respectively: the ratio of the covariances
is scale-dependent with large differences at small scales that
become smaller for wavenumbers $k\gsim 0.1\k$. One might imagine that the difference between the covariances can be modeled such that the true covariance can be recovered through a rescaling of the covariance from paired-and-fixed simulations. However, this does not seem to be a feasible path forward, not only because of the complicated scale- and redshift-dependence of the differences, but also because the paired-and-fixed $C(k,k^\prime)$ is
much noisier than the Gaussian simulations. In order to reach
the same level of noise one would need to make many more realisations of paired-and-fixed simulations,
which seems to defeat the motivation for the paired-and-fixed
simulations in the first place, i.e., to reduce dramatically the need for many realisations.

\makeatletter{}\section{Bispectrum}
\label{sec:bispectrum}
The bispectrum is defined as a product of amplitudes of Fourier
harmonics at three wavenumbers:
$B(k_1,k_2,k_3)=\langle \delta_{k_1}\delta_{k_2}\delta_{k_3}\rangle$.  
It is convenient to scale out the main dependence on the power spectrum, and define the reduced bispectrum as
\begin{align}
B_{reduced} &= \frac{\langle\delta_{k_1}\delta_{k_2}\delta_{k_3}\rangle}{P(k_1)P(k_2)+P(k_2)P(k_3)+P(k_1)P(k_3)},
\end{align}
where the $P(k_i)$ in the denominator are the average power spectra.

\begin{figure} \centering
\includegraphics[width=0.475\textwidth]{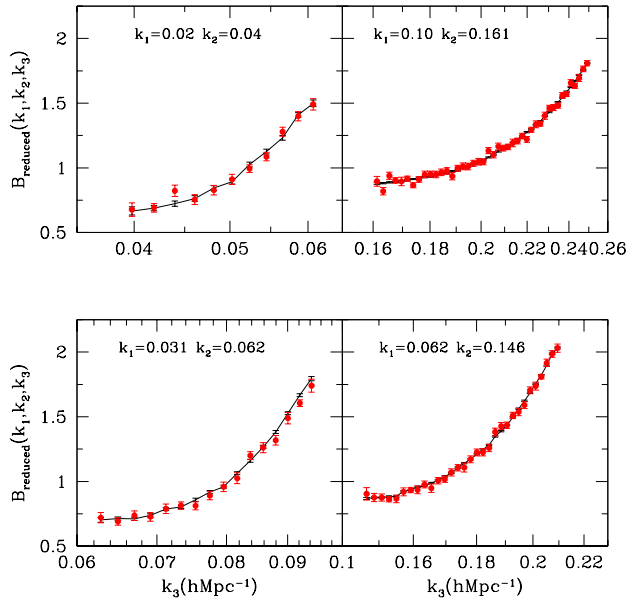}

\caption{Average reduced bispectrum of dark matter at $z=0$ for four slices
  with values of fixed $k_1$ and $k_2$ as indicated in the panels.  Full curves
  show the results of 1300 realisations of Gaussian simulations
  (3GpcGauss). Points with error bars are the results from the combined 3GpcFixed
  and 3GpcPairedFixed simulations (a total of 50 realisations). There are
  no statistically significant differences between true Gaussian and
  paired-and-fixed simulations.}
\label{fig:BiAverage}
\end{figure}

Four slices of the average reduced bispectrum at $z=0$ are presented
in Figure~\ref{fig:BiAverage} for the set of $3\,\Gpch$ simulations (see Table~\ref{table:simtable}). 
Each slice assumes a fixed combination of $k_1$ and $k_2$ while $k_3$ is allowed to vary.
The two panels
on the left show the average bispectrum for small wavenumbers, while the right panels focus on shorter wavelengths around the BAO scale.
Just as we observed for the average power spectrum, we do not find any differences
between Gaussian and paired-and-fixed simulations in the wavenumber domain
discussed here.

\begin{figure}
\centering
\includegraphics[width=0.475\textwidth]{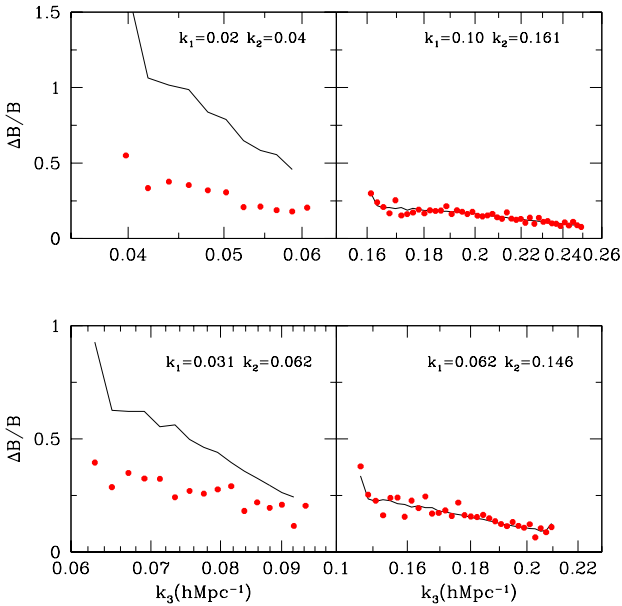}
\caption{The r.m.s. fluctuations of the reduced bispectrum for a single 
  realisation relative to the average bispectrum for the same slices
  as in Figure~\ref{fig:BiAverage}. At very long wavelengths the noise in the 
  bispectrum is a factor of three
  lower for the paired-and-fixed simulation (top-left panel). The difference becomes
  smaller for even slightly shorter wavelengths (bottom-left panel) and
  disappear entirely in the domain of BAOs,
  $k=(0.07-0.30)\,h{\rm Mpc}^{-1}$ (top-right and bottom-right panels).}
\label{fig:BispecRMS}
\end{figure}

Figure~\ref{fig:BispecRMS} shows the scatter of the bispectrum represented by the diagonal components of the covariance matrix. The results follow the
same trend as for the diagonal components of the power spectrum covariance matrix. The scatter is significantly suppressed for very
long wavelengths with $k_1=0.02\k$ and $k_2=0.04\k$ (upper left panel), and it decreases at shorter
wavelengths (bottom left panel). At larger wavenumbers (right panels), the differences disappear. We note that this particular feature is unlike the power spectrum covariance, for which the paired-and-fixed simulations always had a scatter that was suppressed relative to the Gaussian simulations.

  \makeatletter{}

\makeatletter{}

\makeatletter{}\section{Conclusions}
\label{sec:conclusion}
Making accurate theoretical predictions for the clustering statistics
of large-scale galaxy surveys is a very relevant but challenging
process \citep[e.g.,][]{Mandelbaum2013,Sergio2016,DES2018,Uitert2018}.  
One of the main goals is to estimate the
scatter and covariance errors of measured clustering properties 
such as the power spectrum and bispectrum: one needs to know
the covariances in order to estimate the uncertainties of the inferred cosmological
parameters. However, computing the covariance matrices with 
sufficient accuracy typically requires thousands of realisations, which makes the process quite computationally expensive. 

There are tools to create realisations quickly. These include a new 
generation of faster $N$-body codes \citep{Tassev2015,Koda2016,GLAM} and 
approximate methods \citep{Patchy2016,Monaco2016,EZMOCKS2018,Lippich2019}. An alternative approach proposed by \cite{AnguloPontzen} has recently attracted attention: the amplitudes 
of the Fourier harmonics are fixed to be exactly equal to the average amplitude expected 
for a given cosmological model with true Gaussian fluctuations. Thus, in the linear regime, 
even a single realisation with fixed harmonics will have exactly the same power 
spectrum as the average power spectrum obtained from many realisations of true 
Gaussian-distributed harmonics. 
A further reduction in the scatter is achieved by running fixed simulations in pairs, where the simulations in a pair have harmonics with opposite phases.

The density field generated by the fixed 
amplitude method is still a random field: the amplitudes of the Fourier harmonics
are random numbers with values equal to either one or minus one (see eq. (\ref{eq:delta3})).
As a result, there will be some randomness of fluctuations, but it is strongly suppressed and develops 
mainly due to non-linear interactions \citep{PairedStatistics}. Thus,  paired-and-fixed simulations  
seem to be a promising direction to reduce scatter due to cosmic variance and, crucially, to reduce 
the number of realisations for estimates of different statistics. However, there are 
still remaining questions and concerns about the effect of fixing the amplitudes 
of the harmonics. What happens in the non-linear regime?
If the realisation-to-realisation  scatter is suppressed, at least in the linear regime, 
and the scatter is only the diagonal component of the covariance matrices, how are the full covariances affected 
by fixing the amplitudes of the Fourier harmonics? 

There is a clear answer to one of these questions. {\it Every  average} clustering statistic
studied in this paper, and in other works, is not affected in any measurable way by 
paired-and-fixed simulations as compared to those obtained from true Gaussian perturbations.
The list of  statistics is long and includes: the power 
spectrum, halo mass 
function, correlation function,
density distribution function and 
bispectrum \citep[this paper,][]{AnguloPontzen,PairedStatistics,UNIT}.
It is very promising for the paired-and-fixed method that there are no defects, at least for the average statistics.

This is despite the fact that paired-and-fixed fluctuations are clearly not what one expects for initial conditions in the
real Universe. The real fluctuations generated during inflation should not have fixed amplitudes, so  why do paired-and-fixed simulations give the correct answers for 
so many different average statistics? The apparent success of the method is related 
to the fact that the statistics of interest are typically large sums of many 
contributions from different Fourier harmonics. Then, by the virtue of the central  
limit theorem, the distribution function 
of these statistics is as Gaussian as that from truly Gaussian fields in inflationary models.
As an example, we can consider the density field $\delta(\vec x)$. It is a sum of many  
harmonics, as shown in eqs.~(\ref{eq:delta}--\ref{eq:delta4}). Thus, the 
distribution function of $\delta(\vec x)$ and its two-point correlation function 
$\langle\delta(\vec x)\delta(\vec x+\vec r)\rangle$ are the same as for 
a true Gaussian density field. Because the initial density field, together with the adopted
cosmological parameters, define how the different structures evolve and collapse 
in the expanding Universe, it is not surprising that many average properties 
(e.g., correlations functions or halo mass functions) are correctly recovered.

How much of a reduction in the realisation-to-realisation scatter that one gains
by performing paired-and-fixed simulations
depends on the degree of non-linear growth, and consequently on the physical scale. 
For example, at $k=0.05\k$ the r.m.s. fluctuation of the power spectrum is ten times
smaller in paired-and-fixed simulations (see Figure~\ref{fig:DiagCovMatrix}). Because 
the error of the average power spectra in Gaussian simulations declines as $1/\sqrt{N_r}$,
where $N_r$ is the number of realisations, one would need 100 times more Gaussian 
realisations to reach the same accuracy as the average power spectrum from 
paired-and-fixed simulations. A large reduction of noise is similarly observed for the
bispectrum: the r.m.s. fluctuations are reduced by a factor of three 
for $(k_1,k_2,k_3)=(0.02,0.04,0.04)\k$ (see Figure~\ref{fig:BispecRMS}). 

Unfortunately, these gains in the number of realisations diminish dramatically with scale. In the BAO spectral domain $k=0.07-0.3\,h{\rm Mpc}^{-1}$, the r.m.s. fluctuations of the power spectrum are smaller in the 
paired-and-fixed simulations by only  $\sim 20$ per cent compared to the purely 
Gaussian realisations. The r.m.s. scatter in the bispectrum shows an 
even smaller effect: it is the same as
in Gaussian simulations for $k\gsim 0.1\k$ (right panels in Figure~\ref{fig:BispecRMS}).
Similar results are found for those statistics that are more sensitive to strongly non-linear 
interactions. Paired-and-fixed simulations offer no advantage
for the halo mass function \citep{PairedStatistics}, and we do not see  any gains for 
the density distribution function (see Figure~\ref{fig:PDF}).

One of the main goals and challenges for any simulation method is to accurately capture the scatter and covariances of clustering statistics, but in this work we find that this poses a difficult issue for paired-and-fixed simulations. As Figure~\ref{fig:DiagCovMatrix} 
shows, even the scatter of the power spectrum (the diagonal component of the covariance matrix) presents a problem, because the difference in the scatter with respect to true Gaussian simulations depends on the wavenumber $k$ in a complicated way.
One might consider fitting these differences with an analytical approximation in order to recover the true scatter. However, this would not work because that difference 
depends on redshift and must be done for the distribution of mock galaxies, that
in turn requires numerous realisations to be estimated accurately enough.

The non-diagonal components of the power spectrum covariance matrix (Figures~\ref{fig:CovMatrix} 
and \ref{fig:CovMatrixz}) point to another complication for the paired-and-fixed simulations. 
Just as for the scatter of the power spectrum, the ratio of non-diagonal components 
of paired-and-fixed to Gaussian simulations depends on scale and redshift in a 
non-trivial way. In addition, the covariance matrix of paired-and-fixed simulations is much noisier than for purely Gaussian simulations with the same number of realisations. 
In order to reach the same level of noise one would need to make many more 
paired-and-fixed realisations, which defeats the main motivation behind this method.

\medskip
The main results of our work are summarised as follows: 
\begin{itemize}
\item Paired-and-fixed simulations accurately reproduce the average of the power spectrum, PDF and bispectrum from Gaussian simulations.
\item The reduction in cosmic variance error obtained from paired-and-fixed 
simulations is limited to very long-wavelengths with $k\lsim 0.05\k$. Non-linear 
effects erase most of the advantage from pairing and fixing, even in the weakly non-linear domain $k\gsim 0.1\k$.
\item Paired-and-fixed simulations fail to provide a path towards reproducing the true Gaussian covariance errors. 
Not only are the reported differences with Gaussian simulations scale- and redshift-dependent, but covariances from paired-and-fixed simulations 
are also much noisier and would require many more realisations to reach the same level of accuracy 
as obtained from Gaussian simulations.
\end{itemize}

Because accurate covariance errors are crucial for robustly translating the observed clustering statistics from large galaxy surveys into constraints on cosmological parameters, the results presented in this paper demonstrate that paired-and-fixed simulations are not suitable 
for generating mock galaxy catalogs, but they are a very useful tool for quickly and accurately estimating the average properties of the clustering signal.

\section*{Acknowledgements}
The authors thank Raul Angulo and Andrew Pontzen for useful comments and suggestions.
A.K. and F.P. acknowledge support from the State Agency for Research of 
the Spanish MCIU through the grant AYA2014-60641-C2-1-P. J.B. acknowledges support from the Swiss NSF. The GLAM simulations have
been performed on the MareNostrum4 supercomputer at the Barcelona Supercomputer 
Center in Spain.

\bibliography{PMP}
\bibliographystyle{mn2e}

\end{document}